\documentstyle[aps,prb,multicol,psfig]{revtex}
\tighten
\begin{document}
\title{Hall effect in the cuprates: the role of
forward scattering on impurities}
\author{R. Hlubina}
\address{Center for Theoretical Studies, 
ETH H\"onggerberg, CH-8093 Z\"urich, Switzerland\\
Department of Solid State Physics, Comenius University,
Mlynsk\'a Dolina F2, SK-842 48 Bratislava, Slovakia}
\maketitle
\begin{abstract}
We solve the Boltzmann equation for electrons moving in a two-dimensional
plane of square symmetry in the presence of a transverse magnetic field $B$.
We assume that there are two sources of scattering: a large
momentum-independent scattering on a collective mode of the electron system,
and a smaller momentum-dependent forward scattering on impurities. We show
that the effect of impurities on the longitudinal and Hall conductivities is
of the same order of magnitude.
\end{abstract}
\begin{multicols}{2}

Very recently, an interesting proposal has been advanced with the aim of
explaining the anomalous magnetotransport data \cite{Cooper96} in the
cuprates.  Namely, it has been suggested that the marginal Fermi liquid theory
(MFL) which correctly predicts the temperature dependence of the resistivity
of the optimally doped cuprates, $\rho\propto T$, can be modified by taking
into account the scattering on impurities away from the CuO$_2$
planes.\cite{Varma00} It has been argued previously that such impurity
scattering should be of very special type, allowing the electron to change its
momentum by only a small fraction of the Fermi momentum.\cite{Abrahams00} This
peculiar type of scattering was argued in Ref.~\onlinecite{Varma00} to lead to
corrections to the Hall conductivity of the pure MFL system, $\sigma_H\propto
T^{-2}$, which are in agreement with the experimentally observed scaling
$\sigma_H\propto T^{-3}$. In this Report we explore this idea in more
detail. In particular, we ask whether within such modified MFL approach, both
the resistivity and the Hall conductivity data can be explained on equal
footing.

Let us first introduce the model under study.  We consider electrons moving in
a two-dimensional plane of square symmetry. We assume that the Fermi sea is a
simply connected region in ${\bf k}$ space whose boundary (the Fermi line) has
a length $2\pi k_F$.  We shall numerate the points on the Fermi line by a
dimensionless length $\varphi$ defined by $d\varphi=dk/k_F$, where $dk$ is an
element of the Fermi line. We set $\varphi=0$ along the $x$ axis of the plane
(which is assumed to coincide with one of the crystallographic axes of the
plane).

Within standard transport theory,\cite{Ziman60} we want to study the transport
properties of the electron gas in an applied electric field ${\bf E}$ parallel
to the plane and a magnetic field ${\bf B}$ perpendicular to the
plane. Because of the square symmetry of the problem, the linear response
coefficients do not depend on the direction of ${\bf E}$ and we take it to be
parallel to the $x$ axis of the plane.  Let the local (in ${\bf k}$-space)
departure of the distribution function of the electrons, $f_{\bf k}$, from the
equilibrium distribution, $f^0_{\bf k}$, be $f_{\bf k}=f_{\bf
k}^0-(eE/k_F)g_{\bf k}\partial f_{\bf k}^0/\partial\varepsilon_{\bf k}$, where
$\varepsilon_{\bf k}$ is the quasiparticle energy.

Let us introduce the local Fermi energy $\varepsilon_F({\bf k})=v_{\bf
k}K_{\bf k}$, where $v_{\bf k}$ is the local Fermi velocity and $K_{\bf k}$ is
the local radius of curvature of the Fermi line. Then for temperatures $T\ll
{\rm min}[\varepsilon_F({\bf k})]$ the Boltzmann equation for quasielastic
scattering on bosonic excitations and impurities reads\cite{Hlubina00}
\begin{equation}
\cos\psi(\varphi)+\beta g^\prime(\varphi)
=\oint{d\varphi^\prime\over 2\pi}A(\varphi,\varphi^\prime)
\left[g(\varphi)-g(\varphi^\prime)\right],
\label{eq:Boltzmann}
\end{equation}
where $A(\varphi,\varphi^\prime)$ describes the scattering of the electrons
between the points $\varphi$ and $\varphi^\prime$ of the Fermi surface and
$\beta=eB/\hbar k_F^2$ is a dimensionless magnetic field.  $\psi(\varphi)$ is
the angle between the normal to the Fermi line in the point $\varphi$ and the
$x$ direction, $\cos\psi={\bf E}\cdot{\bf v}_{\bf k}/Ev_{\bf k}$. Note that
Eq.~(\ref{eq:Boltzmann}) is valid for a general shape of the Fermi surface
with a non-constant density of states (and, thus, a non-constant $v_{\bf k}$)
along the Fermi line. The information about $v_{\bf k}$ is contained in the
dimensionless scattering function $A(\varphi,\varphi^\prime)$.
\cite{Hlubina00}

In Ref.~\onlinecite{Varma00}, the following scattering function
has been proposed to describe the magnetotransport in the cuprates:
$$
A(\varphi,\varphi^\prime)=\Gamma_1+
A_2\left({\varphi+\varphi^\prime\over 2},\varphi^\prime-\varphi\right),
$$
where $\Gamma_1$ describes the scattering of the electrons on a hypothesized
MFL mode. Since this scattering is supposed to be momentum-independent,
$\Gamma_1$ is a weak function of $\varphi,\varphi^\prime$, which we model by a
constant.  The physical (dimensionful) electron lifetime $\tau_{\bf k}$ is
related to $\Gamma_1$ via $\tau^{-1}_{\bf k}=v_{\bf k}k_F\Gamma_1$.  Within
the MFL phenomenology, it is assumed that $\Gamma_1$ exhibits an anomalous
scaling with temperature, $\Gamma_1\propto T$.  

The new ingredient introduced in Ref.~\onlinecite{Varma00} is the scattering
on impurities which is described by the function $A_2(\varphi,\theta)$. The
authors of Ref.~\onlinecite{Varma00} argue rather convincingly that this type
of scattering is effective only for $\theta<\theta_c$, i.e., the scattering
is in the forward direction.  

Since $\theta_c\sim(k_Fd)^{-1}$ where $d$ is the characteristic distance of
the impurities from the CuO$_2$ plane, the actual numerical value of
$\theta_c$ may be not too small. Therefore, in what follows we shall consider
two limiting cases: $\theta_c\ll 1$ and $\theta_c\sim 2\pi$.  We show that in
both limits the impurity contribution leads to effects of the same order of
magnitude, when expressed in terms of the impurity transport lifetime.

{\it Forward scattering on impurities.} In this case $\theta_c\ll 1$ and, as
shown in Ref.~\onlinecite{Hlubina00}, it is useful to define a (dimensionless)
transport scattering rate
$$
\Gamma_2(\varphi)=\oint {d\theta\over 2\pi}
A_2(\varphi,\theta)(1-\cos\theta).
$$
Note that the scattering rate $\Gamma_2(\varphi)$ is not assumed to be
constant. Rather, the authors of Ref.~\onlinecite{Varma00} suggest that
$\Gamma_2(\varphi)$ should be large along the Cu-O-Cu bonds and small along
the $(\pm\pi,\pm\pi)$ directions, in order to make the single particle
lifetime compatible with the lifetime anisotropy deduced from the ARPES
experiments on the cuprates.\cite{Abrahams00}

Making use of the scattering rate $\Gamma_2(\varphi)$, the Boltzmann equation
(\ref{eq:Boltzmann}) simplifies to
\begin{equation}
\cos\psi+\beta g^\prime=
\Gamma_1g-\left(\Gamma_2g^\prime\right)^\prime,
\label{eq:f_Boltzmann}
\end{equation}
where the primes denote derivatives with respect to $\varphi$.  In
Eq.~(\ref{eq:f_Boltzmann}), scattering on MFL fluctuations is treated in the
relaxation-time approximation, whereas scattering on impurities is described
within the recently developed scheme for forward scattering.\cite{Hlubina00}

We assume that $\beta\ll 1$ and we expand $g$ in powers of $\beta$ to first
order in $\beta$, $g=g_0+g_1$, where $g_n\propto\beta^n$.  We assume
furthermore that $\Gamma_1\gg\Gamma_2$ for $T>100$ K and we calculate $g_0$
and $g_1$ to the lowest nontrivial order in ${\tilde\tau}_1\Gamma_2$, where
${\tilde\tau}_1=\Gamma_1^{-1}$.  These assumptions are checked at the end of
the calculation, when we compare our results to the experimental data on the
cuprates.  With the above simplifications, we find
\begin{eqnarray}
g_0&=&{\tilde\tau}_1\left[h+{\tilde\tau}_1\left(\Gamma_2
h^\prime\right)^\prime\right],
\label{eq:g0}
\\
g_1&=&\beta{\tilde\tau}_1^2
\left[h^\prime+{\tilde\tau}_1\left(\Gamma_2 h^\prime\right)^{\prime\prime}
+{\tilde\tau}_1\left(\Gamma_2 h^{\prime\prime}\right)^\prime\right],
\label{eq:g1}
\end{eqnarray}
where we have introduced an auxiliary function $h(\varphi)=\cos\psi$. In
agreement with Ref.~\onlinecite{Varma00}, a term proportional to the second
derivative of $\Gamma_2$ appears in Eq.~(\ref{eq:g1}).

Following Ref.~\onlinecite{Hlubina00}, we calculate the
longitudinal and Hall conductivities, respectively, as follows:
\begin{eqnarray}
\sigma&=&{2e^2\over h}\oint {d\varphi\over 2\pi}
g_0(\varphi)\cos\psi(\varphi),
\\
\sigma_H&=&-{2e^2\over h}\oint {d\varphi\over 2\pi}
g_1(\varphi)\sin\psi(\varphi).
\end{eqnarray}
In taking the integrals, we repeatedly make use of the trigonometric
relations $\cos^2\psi=(1+\cos 2\psi)/2$ and 
$\sin^2\psi=(1-\cos 2\psi)/2$ and of the identity 
$$
\oint {d\varphi\over 2\pi}\cos2\psi F(\varphi)=
\oint {d\varphi\over 2\pi}\sin2\psi F(\varphi)=0,
$$
which holds for any function $F(\varphi)$ compatible with square symmetry.  In
fact, under the transformation $\varphi\rightarrow \varphi+\pi/2$,
$F(\varphi)$ does not change, whereas
$\psi(\varphi+\pi/2)=\psi(\varphi)+\pi/2$ and hence $\cos 2\psi$ and $\sin
2\psi$ change sign.

Integrating per parts so as to remove the derivatives of the function
$\Gamma_2$ and making use of the above identities, we find the resistivity
$\rho=\sigma^{-1}$ and the Hall angle $\Theta_H=\sigma_H/\sigma$,
\begin{eqnarray}
\rho&=&{h\over e^2}\left[{1\over{\tilde\tau}_1}+\oint{d\varphi\over 2\pi}
\Gamma_2(\varphi)(\psi^\prime)^2\right],
\label{eq:f_rho}
\\
\Theta_H&=&\beta{\tilde\tau}_1\left[1+{\tilde\tau}_1\oint{d\varphi\over 2\pi}
\Gamma_2(\varphi)(\psi^\prime)^2(1-2\psi^\prime)\right].
\label{eq:f_Hall}
\end{eqnarray}
The factors $\psi^\prime$ are determined by the shape of the Fermi line.  For
a circular Fermi line, $\psi^\prime=1$. For non-circular Fermi lines,
$\psi^\prime$ oscillates around 1, being smaller (larger) in the flat (curved)
parts of the Fermi line. \cite{note1} 

{\it s-wave scattering on impurities.}  If $\theta_c\sim 2\pi$, electrons can
scatter on an impurity to all directions. In this so-called s-wave scattering
case the scattering function
$A_2[(\varphi+\varphi^\prime)/2,\varphi^\prime-\varphi]$ becomes a function of
the incoming momentum only, i.e.  $A_2\rightarrow \Gamma_2(\varphi)$ and the
standard relaxation-time approximation applies both to MFL scattering and to
impurity scattering.  Thus the Boltzmann equation simplifies to
\begin{equation}
\cos\psi+\beta g^\prime=\left[\Gamma_1+\Gamma_2(\varphi)\right]g.
\label{eq:s_Boltzmann}
\end{equation}
Note the difference of this equation with respect to
Eq.~(\ref{eq:f_Boltzmann}).

Assuming again that $\Gamma_1\gg\Gamma_2$ and calculating $g_0$ and $g_1$ to
the lowest nontrivial order in ${\tilde\tau}_1\Gamma_2$, we find results
identical to Eqs.~(\ref{eq:f_rho},\ref{eq:f_Hall}), with the only change that
we should write $\Gamma_2(\varphi)$ instead of
$\Gamma_2(\varphi)(\psi^\prime)^2$.  Thus, if $\psi^\prime\approx 1$ (which is
the case in the cuprates \cite{Randeria97}), the impurity effects are
virtually the same in both limiting cases (provided they are expressed in
terms of the transport scattering rate $\Gamma_2$).

In a previous paper we have shown that magnetotransport is completely
different in systems with dominant forward and s-wave
scattering.\cite{Hlubina00} Thus our present result might come as a
surprise. However, there is nothing mysterious about it.  In the model of
Ref.~\onlinecite{Varma00}, the dominant scattering is on the MFL mode. This
scattering is of s-wave type and as such is well describable by the
relaxation-time approximation.  The impurity scattering is only a small
perturbation which can not manifest itself too differently in the limiting
cases of forward and s-wave scattering. In some sense, this is similar to the
analysis of impurity scattering at low temperatures in nearly
antiferromagnetic systems.\cite{Rosch00} In that case, impurities are the
s-wave scatterer and antiferromagnetic fluctuations are the anomalous
scatterer.  If the s-wave scattering dominates (which happens typically at low
temperatures), then it is not necessary to search for full solutions of the
Boltzmann equation as would be the case in a clean system, \cite{Hlubina95}
and the temperature dependence of the transport coefficients can be 
determined making use of the relaxation-time approximation.

{\it Anisotropic} ${\tilde\tau}_1$.  Let us consider briefly the effect of a
possible anisotropy of ${\tilde\tau}_1$.  After all, within MFL theory one
requires that it is the physical lifetime, $\tau_{\bf k}$, which is isotropic
and thus, if the Fermi velocity is not constant around the Fermi line, then
${\tilde\tau}_{1 {\bf k}}= \tau_{\bf k}v_{\bf k}k_F$ should be anisotropic as
well.  For a non-constant ${\tilde\tau}_1$, the longitudinal and Hall
conductivities read, again to leading non-trivial order in
${\tilde\tau}_1\Gamma_2$,
\begin{eqnarray}
\sigma&=&{e^2\over h}\oint{d\varphi\over 2\pi} 
\left[{\tilde\tau}_1-\Gamma_2{\cal T}_1\right],
\label{eq:rho}
\\
\sigma_H&=&{e^2\over h}\beta\oint{d\varphi\over 2\pi}{\tilde\tau}_1
\left[{\tilde\tau}_1\psi^\prime-2\Gamma_2 {\cal T}_2\right].
\label{eq:Hall}
\end{eqnarray}
For forward impurity scattering we find ${\cal
T}_1={\tilde\tau}_1^2(\psi^\prime)^2 +({\tilde\tau}_1^\prime)^2$ and ${\cal
T}_2={\tilde\tau}_1^2(\psi^\prime)^3
+{\tilde\tau}_1{\tilde\tau}_1^\prime\psi^{\prime\prime}
+2({\tilde\tau}_1^\prime)^2\psi^\prime
-{\tilde\tau}_1{\tilde\tau}_1^{\prime\prime}\psi^\prime$, whereas for s-wave
scattering on impurities ${\cal T}_1={\tilde\tau}_1^2$ and ${\cal
T}_2={\tilde\tau}_1^2\psi^\prime$.

{\it Discussion.} Let us apply the above results to the cuprates.  Taking
$k_F\approx 0.74$ \AA$^{-1}$ we find $\hbar k_F^2/e\approx 3.6\times 10^4$ T,
confirming our assumption that $\beta\ll 1$ for laboratory fields.  In what
follows, we shall assume the simplest nontrivial angular variations of the
quantities ${\tilde\tau}_1(\varphi)$, $\psi(\varphi)$, and
$\Gamma_2(\varphi)$.  The dimensionless MFL lifetime is assumed to vary along
the Fermi line according to ${\tilde\tau}_1(\varphi)=\tau_0(1-\delta\cos
4\varphi)$ with $0<\delta<0.1$, taking into account the slightly smaller
Fermi velocity along the Cu-O-Cu bonds.\cite{Valla00} The shape of the Fermi
line is modelled by $\psi(\varphi)=\varphi-\epsilon\sin 4\varphi$ with
$0<\epsilon<0.25$, in accordance with a flat Fermi line at $\varphi=0$ and
equivalent directions.\cite{Randeria97} Finally, we take
$\Gamma_2(\varphi)=\Gamma_0\cos^2 2\varphi$, as required by the recent ARPES
experiments.\cite{Valla00} Within MFL theory it is assumed that $\tau_0\propto
T^{-1}$ and $\Gamma_0$ is temperature-independent.  Inserting the above
expressions into Eqs.(\ref{eq:rho},\ref{eq:Hall}), we obtain
\begin{eqnarray}
\rho&=&{h\over e^2}
\left[{1\over\tau_0}+\Gamma_0 f_1\right],
\label{eq:rho_fin}
\\
\theta_H&=&\tau_0\beta
\left[1+4\delta\epsilon + \tau_0\Gamma_0 
\left(f_1(1+4\delta\epsilon)+f_2\right)\right].
\label{eq:Hall_fin}
\end{eqnarray}
For forward scattering on impurities we find, to linear order in $\delta$,
\begin{eqnarray*}
f_1(\epsilon,\delta)&=&{1\over 2}
(1-4\epsilon+8\epsilon^2)-
{\delta\over 2}(1 - 8\epsilon + 12\epsilon^2),
\\
f_2(\epsilon,\delta)&=&-(1-6\epsilon+24\epsilon^2-24\epsilon^3)\\
&& +2\delta(5-44\epsilon+36\epsilon^2-48\epsilon^3).
\end{eqnarray*}
For s-wave impurity scattering we obtain (again to linear order in $\delta$)
$f_1(\epsilon,\delta)=(1-\delta)/2$ and $f_2(\epsilon,\delta)=-(1-2\epsilon)+
\delta(3-12\epsilon)/2$.

Turning to the resistivity data, experiment requires that at $T\approx 100$ K,
$\tau_0\Gamma_0f_1\approx 1/9$, since the ratio of the resistivity at 100 K to
its 0 K extrapolated value is $\approx 10$.\cite{Varma00} Note that since
$f_1\approx 1/2$, this justifies a posteriori our assumption
${\tilde\tau}_1\Gamma_2\ll 1$ already at $T=100$ K. At higher temperatures,
${\tilde\tau}_1\Gamma_2$ becomes even smaller.  

As regards the Hall angle, the impurity contribution should dominate at $T>
100$ K, if the mechanism proposed in Ref.\onlinecite{Varma00} has to apply.
This requires $\tau_0\Gamma_0[f_1(1+4\delta\epsilon)+f_2]\gg
1+4\delta\epsilon$, or, taking into account the estimate of
$\tau_0\Gamma_0f_1$ (at $T=100$ K) from the resistivity data, $f_2\gg
8f_1(1+4\delta\epsilon)$. This is, however, impossible for $0<\delta<0.1$ and
$0<\epsilon<0.25$, as can be readily seen both for forward and s-wave impurity
scattering.

{\it Conclusions.} Within standard transport theory we have shown that
additional impurity scattering on top of a dominant isotropic scattering on a
collective mode does indeed lead to corrections to the Hall number, as
predicted in Ref.~\onlinecite{Varma00}. However, the effect is sufficiently
large only for impurity scattering comparable to the inelastic scattering, in
which case also the impurity contribution to the resistivity becomes
comparable to the inelastic (MFL) contribution.  Thus the resistivity and the
Hall number observed experimentally in the cuprates can not be explained
simultaneously within the picture advanced in Ref.~\onlinecite{Varma00}.

{\it Acknowledgement.}
I thank Professor T.~M.~Rice and the Institut f\"ur theoretische Physik, ETH
Z\"urich, for their hospitality. This work was partially supported by the
Slovak Grant Agency VEGA under Grant No. 1/6178/99.

\end{multicols}

\begin{references}
\bibitem{note1}For Fermi surfaces with inflection points (such as those
occuring in the $t$-$t^\prime$ model slightly below the Van Hove density),
$\psi^\prime$ can even change sign.
\end{references}
\end{document}